\input epsf 
\documentstyle[preprint,aps,eqsecnum,tighten]{revtex}
\preprint{\vbox{\baselineskip=12pt
\rightline{CGPG-95/8-}
\rightline{gr-qc/9601011}}}

\begin{document}
\title{
$2+1$ Covariant Lattice Theory and t'Hooft's Formulation
}
\author{
Henri Waelbroeck\thanks{hwael@roxanne.nuclecu.unam.mx}\\ 
{\em Instituto de Ciencias Nucleares, UNAM, Circuito Exterior CU} \\
{\em A Postal 70-543, M\'exico DF 04510, Mexico}\\ 
Jos\'e A. Zapata\thanks{zapata@phys.psu.edu}\\ 
{\em Center for Gravitational Physics and Geometry} \\
{\em Department of Physics, The Pennsylvania State University} \\
{\em 104 Davey Laboratory, University Park, PA 16802, USA}
}

\maketitle

\begin{abstract}
 We show that 't Hooft's representation of (2+1)-dimensional gravity 
in terms of flat polygonal tiles is closely related to a gauge-fixed 
version of the covariant Hamiltonian lattice theory. 't Hooft's gauge is 
remarkable in that it leads to a Hamiltonian which is a linear sum of 
vertex Hamiltonians, each of which is defined modulo $2 \pi$.
A cyclic Hamiltonian implies that ``time'' is quantized. However, it 
turns out that this Hamiltonian is {\it constrained}. If one chooses 
an internal time and solves this constraint for the ``physical Hamiltonian'', 
the result is not a cyclic function. Even if one quantizes {\it a la Dirac}, 
the ``internal time'' observable does not acquire a discrete spectrum. 
We also show that in Euclidean 3-d lattice gravity, ``space'' can be 
either discrete or continuous depending on the choice of quantization.
Finally, we propose a generalization of 't Hooft's gauge for Hamiltonian 
lattice formulations of topological gravity dimension 4.

\noindent
PACS number(s):0240,0420

\end{abstract}
\pacs{PACS number(s):0240,0420}

\section{ Introduction }
 Because (2+1)-dimensional gravity is closely related to the Chern-Simons
topological invariant for the gauge group $ISO(2,1)$, an exact lattice 
version of the theory exists \cite{waelbroeckCLT}. This in turn has led 
to a convenient parametrization of the reduced phase space, the moduli space 
of flat $ISO(2,1)$ connections, known as the ``polygon representation''
of (2+1)--dimensional gravity. In both the lattice theory and the polygon
representation, the translations of the vertices are generated by
first-class constraints, so the gauge is never fixed -- rather, the reduced
variables are independent of the foliation of spacetime. Having thus 
decoupled the physical degrees of freedom from the gauge, the canonical
quantization programme can be carried out explicitly \cite{BwedgeF}, 
leading to a proposal for the modular-invariant wave function, in terms of 
a sum of plane waves analogous to diffraction by a grating \cite{Quebec}.

 Proceeding from a different angle, 't Hooft has proposed a gauge-fixed 
representation of (2+1)--dimensional gravity based on a foliation with a 
patchwork of flat polygons\cite{'thooftCLASS}. With this choice of gauge, 
the Hamiltonian 
becomes a linear sum of the deficit angles at the vertices, rather than 
the usual trace of a product of non-commuting Lorentz matrices. The cyclic 
nature of the Hamiltonian, which is then only defined modulo $2 \pi$, 
indicates that time is quantized \cite{'thooftQUANT}. 

 In this letter, we throw a bridge between the exact lattice theory
and 't Hooft's flat polygons. To do so, we solve the $SO(2,1)$ Gauss 
constraint at the lattice faces by changing to scalar variables. 
The resulting symplectic structure agrees
with that proposed by 't Hooft, and the dynamical evolution is generated
by an arbitrary linear combination of first-class translation constraints. 
On a tessellation in terms of flat polygons, these constraints reduce to 
conditions on the deficit angles; the sum of which is the Hamiltonian 
proposed by 't Hooft. 

 The canonical gauge fixing procedure, however, would require solving the 
constraints together with the gauge conditions, or alternatively 
computing Dirac brackets which cancel the fluctuations of the gauge 
variables. This would lead to non-local Dirac brackets for the scalar 
variables, so it seems preferable to invoke instead the relation between 
't Hooft's scalar variables and the covariant lattice theory. Carrying on with 
the canonical gauge fixing, one would choose an ``internal time'' to compute the
physical Hamiltonian, by solving the Hamiltonian constraint. Although this
constraint is a cyclical function, its solution for the momentum canonically
conjugate to the internal time is not. Therefore, even if one considers time as 
an observable and quantizes {\it a la Dirac}, there is no prediction of a 
discrete spectrum. Rather, the original argument for the discretization 
of time appears to be an artifice of the procedure whereby the Hamiltonian 
was postulated by demanding that it generate the correct dynamics. 

 One only has quantization of time if one does not fix the gauge,
but nonetheless uses an initial condition which satisfies the gauge conditions 
and notes that the Hamiltonian preserves these conditions -- seen from
this perspective, it appears that 't Hooft's proposal is not a canonical
quantization of a diffeomorphism-invariant gravity theory, but the
quantization of a lattice model which represents the same classical solutions --
the conclusion, that classically equivalent formalisms differ as quantum
theories,
should come as no surprise.

 If one considers the gauge group $ISO(3)$ rather than $ISO(2,1)$, both the 
exact lattice theory and the polygon representation predict a discrete spectrum 
for the links, as these are represented by rotation operators. In contrast, if
one follows 't Hooft's procedure one finds that the cyclic form of the 
spatial translation generators imply the discretization of the link lengths
as $l_i = k$, rather than $l_i^2 = k(k+1)$. Finally, in the scalar theory 
which results from fixing the SO(3) symmetry and the translation symmetry 
{\it before} quantizing, the lattice link lengths have a continuous spectrum. 

 These results indicate that the prediction of a discrete structure for
space and time can depend on the choice of quantization, particularly on 
whether one quantizes before or after fixing the gauge.

\section{From the covariant lattice to 't Hooft's polygons}
Now we briefly review the covariant lattice theory
\cite{waelbroeckCLT} based on a lattice version of the Ashtekar-Witten
variables \cite{ashtekar,witten}. To each lattice face we assign a
reference frame which we denote by an index $(i,j, \ldots )$. The role
of the connection is played by three-dimensional Lorentz matrices
${\bf M_{ij}}$ that define parallel transport from face $(j)$ to face
$(i)$. If face $(i)$ is surrounded by faces $(j) , (k) , (l) \ldots $,
the boundary of face $(i)$ is described {\em in the frame assigned to
$(i)$} by 3-vectors ${\bf E}_{ij} , {\bf E}_{ik} , {\bf E}_{il} \ldots $  
(figure $1$). The following identities hold (we will use parentheses
for the lattice indices to avoid confusion with the frame indices).    

\nopagebreak[3]\begin{eqnarray}
&&E_{(ij)}^{\ \ \ a}=- M_{(ij)\ b}^{\ \ \ a} E_{(ji)}^{\ \ \ b} 
\label{anti}\\
&&M_{(ij)\ c}^{\ \ \ a}  M_{(ji)\ b}^{\ \ \ c}= \delta ^a_{\ b} 
\label{inv}\\
&&M_{(ij)\ c}^{\ \ \ a} M_{(ij)}^{\ \ \ bc} = \eta ^{ab} \quad .
\label{orto}
\end{eqnarray}

\noindent 
The symplectic structure is defined by the Poisson brackets 

\begin{eqnarray}
\{E_{(ij)}^{\ \ \ a} , E_{(ij)}^{\ \ \ b} \} &=& \varepsilon ^{ab}_
{\ \ d}  E_{(ij)}^{\ \ \ d} 
\label{4}
\\
\{E_{(ij)}^{\ \ \ a} , M_{(ij)\ c}^{\ \ \ b} \} &=& 
\varepsilon ^{ab}_
{\ \ d}  M_{(ij)\ c}^{\ \ \ d} 
\label{5}
\\
\{E_{(ij)}^{\ \ \ a} , M_{(ji)\ c}^{\ \ \ b} \} &=&- \varepsilon 
^{ad}_{\ \ c}  M_{(ji)\ d}^{\ \ \ b} \quad ,
\label{6}
\end{eqnarray}

\noindent 
which were originally derived from a Chern-Simons action
\cite{waelbroeckCLT} (all other brackets are null).  

%
\noindent 
{\small {\bf Fig. 1}A triangular lattice. 
The labels $i, j, k$ etc denote faces 
of the lattice. The vector ${\bf E}_{ji}$, in the frame which is associated 
to face $j$, represents the boundary between faces $j$ and $i$; this same 
vector points from the vertex $J$ to the vertex $I$.
} 

We require that each face $(i)$ closes, that the curvature vanishes at vertices
$(I)$ with no particles and that spacetime has a conical singularity of deficit
angle $2 \pi m_p$ at vertices $(p)$ with particles 
of mass $m_p$

\begin{eqnarray}
J_{(i)}^{\ \ a} &=&E_{(ij)}^{\ \ \ a}+E_{(ik)}^{\ \ \ a}+\ldots 
\approx 0 
\label{cerra}
\\
W_{(I)\ c}^{\ \ b}&=&({\bf M}_{ij}{\bf M}_{jk}\ldots {\bf M}_{ni})^b_{\ c} 
\approx  \delta ^b_{\ c} 
\label{plani}
\\
\hbox{tr}({\bf W}_{(p)}) &\approx &1 + 2 \cos(m_p) 
\label{trw} \quad ,
\end{eqnarray}

\noindent 
where the vertex $(I)$ is shared by faces $(i) , (j) , \ldots
(n)$. The constraints (\ref{plani}), (\ref{trw}) can be replaced by  
$P_{(I)}^a := \frac{1}{2} \varepsilon _{\ b}^{a\ c} W^{\ \ b}_{(I)\ c} 
\approx 0$ and 
${\bf P}^2(p) -  m_p^2  \approx 0$. These constraints are  first-class 
and generate Lorentz transformations of the frame at $(i)$, translations 
of vertex $(I)$, and time reparametrizations of particle $(p)$'s world 
line  respectively:  

\begin{eqnarray}
&&\{J_{(i)}^{\ \ a} , E_{(ij)}^{\ \ \ b} \} = \varepsilon ^{ab}_
{\ \ d}  E_{(ij)}^{\ \ \ d} \\
&&\{J_{(i)}^{\ \ a} , M_{(ij)\ c}^{\ \ \ b} \} = \varepsilon ^{ab}_
{\ \ d}  M_{(ij)\ c}^{\ \ \ d} \\
&&\{\xi ^a P_{(I)\ a} , E_{(ij)}^{\ \ \ b} \} 
\approx \xi ^b \\
&&\{P_{(p)}^a P_{(p)\ a} , E_{(ij)}^{\ \ \ b} \} 
\approx P_{(p)}^b \quad .
\end{eqnarray}

\noindent 
(The weak equality indicates that the constraints have been used in
simplifying the last brackets.)  

Once we know the variables, restrictions, and symmetries we can
compute the dimension of the physical phase space. The lattice has $N$ particles,

$N_0- N$ vertices with no particles, $N_1$ links, and $N_2$ faces. There are
$6N_1$ phase
space variables ${\bf E_{ij}}$ and ${\bf M_{ij}}$, and $3(N_0-N+N_2)+N$
constraints and symmetries, so that the dimension of the phase space
is  
\nopagebreak[3]\begin{eqnarray} 
6N_1 -2 (3N_0 + 3N_2) +4N= 4N-6\chi =4N+12 g -12
\end{eqnarray} 
\noindent 
where $\chi$ is the Euler number and $g$ the genus of the surface. 

Now we solve the Gauss constraint (\ref{cerra}) by changing to scalar
variables. For the triangular lattice, a good set of variables is the
link lengths $l_{(ij)}$ and the (hyperbolical)angles $\eta _{(ij)}$ between
neighboring faces  
\nopagebreak[3]\begin{eqnarray} 
l_{(ij)}^2 &:=& E_{(ij)a} E_{(ij)}^a \\
\cosh(\eta _{(ij)}) &:=& \frac{1}{N_{(i)} N_{(j)}}N_{(i)}^a (M_{(ij)}N_{(j)})_a
\quad .
\end{eqnarray} 
\noindent 
where $N_{(i)}^a= \varepsilon ^{abc} E_{(ij)a} E_{(ik)b}$.  These $2N_1$
variables
encode all the scalar information given in the $6N_1$ covariant
variables ${\bf E}_{(ij)}$, ${\bf M}_{(ij)}$. 
For a triangular lattice the number of scalar variables ($2N_1=3N_2$)
equals the number of covariant variables minus constraints and
symmetries ($6N_1 -3N_2 - 3N_2 = 2N_1 $). Thus, the dimension of the phase
space after reducing the Lorentz gauge freedom is $2 N_1$.  We could
check from the geometry of the lattice that the scalar variables are
independent. Instead, we see that $l_{(ij)}$ and $\eta _{(ij)}$ are
canonically conjugated. Their independence in the reduced phase-space
is then guaranteed. 
 
After reducing the phase space by gauge fixing and solving the Gauss
law, the relevant brackets of the theory are the Dirac brackets. 
However, an immediate consequence of working with scalar variables
is that their Dirac brackets coincide with their Poisson brackets. By
definition, if a pair of functions $f ,g$ is scalar $\{f,J(i)^a\} = \{
g, J(i)^a \} =0 $ for all $(i)$, then  
  
\nopagebreak[3]\begin{eqnarray} 
\{ f , g \}_{DB} &=& \{ f , g \} - \{ f , G_A \} M_D^{-1\ AB} \{ G_B , g \} 
\nonumber \\ 
&=& \{ f , g \}
\end{eqnarray} 

\noindent 
where some gauge fixing conditions and the gauge generators 
$J(i)^a \approx 0$ were written in a collective fashion as $G_A\approx 0$, and 
$M_{D\ AB}:= \{ G_A , G_B \}$. 

The problem is now to find the brackets of the scalar variables. As an
intermediate step we calculate $ \{ E_{(ij)}^b , \cosh(\eta _{(ij)})\}$ by
calculating the brackets of ${\bf E}_{(ij)}$ with both sides of 
$N_{(i)}^a (M_{(ij)}N_{(j)})_a = N_{(i)} N_{(j)} \cosh(\eta _{(ij)})$; our result
is           

\nopagebreak[3]\begin{eqnarray} 
\{ E_{(ij)}^b , \cosh(\eta _{(ij)}) \} &=&
\frac{1}{N_{(i)} N_{(j)}} N_{(i)}^a \varepsilon _a^{\ cb} 
(M_{(ij)}N_{(j)})_c \nonumber \\ 
&=& \frac{1}{l_{(ij)}} E_{(ij)}^b \sinh(\eta _{(ij)}) \quad , 
\end{eqnarray} 

\noindent 
which has as consequence 
\nopagebreak[3]\begin{eqnarray} 
\{ l_{(ij)}, \eta _{(ij)} \} = 1 \quad . 
\end{eqnarray} 

The curvature constraints (\ref{plani}), (\ref{trw}) induce constraints 
on the scalar phase space spanned by $l_{ij}, \eta _{(ij)}$. These are
first-class constraints that generate translations of the vertices; however, 
in terms of the scalar variables, the translation generators are generally
complicated non-local expressions. To determine the explicit form of 
the induced constraints,
we go back to the covariant description by choosing a gauge. We choose 
auxiliary reference frames at all the faces to embed the
vectors ${\bf E_{ij}}$, with the restriction that they form closed
triangles with edge lengths $l_{ij}$. Once the vectors ${\bf E}$ 
are set in local frames, we prescribe the relative orientation between
neighboring frames with the parallel transport matrices ${\bf
M}$. These matrices are  subject to two requirements, e.g. in the case
of the matrix ${\bf M}_{ij}$, the hyperbolical angle between face
$(i)$ and face $(j)$ has to be $\eta _{(ij)}$ 
and the identity ${\bf E_{(ij)} =-M_{(ij)} E_{(ji)}}$ must hold.  

To recover a simple form of the curvature constraints, we have to choose 
a gauge that is well-tailored for the scalar variables. 
In regions of the lattice where the extrinsic curvature vanishes,
i.e., where $\eta _{(ij)} =0$, we know that the parallel transport around
a vertex, with no particle siting on it, is given by the matrix  

\nopagebreak[3]\begin{eqnarray} 
W(v)^a_{\ b} = \exp(\alpha _v \hat{N}_{(v)}^c \varepsilon _{c\ b}^{\ a}) 
\label{w=expn} 
\end{eqnarray} 
\noindent 
where $\hat{N}_{(v)}$ is the unit normal to any of the faces containing $(v)$, 
and $\alpha _v$ is the deficit angle at $v$ 
\nopagebreak[3]\begin{eqnarray} 
\alpha _v [l's] =2 \pi -{\sum _i} \alpha _i \approx 
\alpha _v^* [\eta 's] = 0 
\label{a=0} \quad .
\end{eqnarray} 

\noindent 
In the zero extrinsic curvature case, the only part of the parallel
transport matrix that is not automatically equal to the identity
matrix is the rotational part. The weak equality (\ref{a=0}), that
sets the deficit angle $\alpha (v)$ to zero holds only for the zero
extrinsic curvature case, and is what the generator of 
translations of $(v)$ in the direction normal to the
lattice. Obviously, orthogonal translations of all vertices in an
extrinsically flat region of the lattice leave the links' lengths
invariant. 

Zero extrinsic curvature and normal translations produce trivial
dynamics; therefore, the strategy is to choose a gauge in which the
lattice is extrinsically flat in as many links as possible. 
Following 't Hooft, we consider a slicing of spacetime with a lattice 
that enjoys the following properties: 

\begin{enumerate} 
\item Zero extrinsic curvature {\em polygons} cover the lattice.  
The parallel transport  between two neighboring cells 
maps the normal to the normal if their boundary belongs to one of the
polygons, i.e. $\eta _{(ij)}=0$ if the link $(ij)$ is in one polygon. 
\item The boundaries between polygon (that are the union of several
lattice links) are straight lines called {\em bones}; 
bones meet at vertices named {\em joints}. Exactly three bones
meet at each joint.  
\item Particles are located at vertices inside the polygons and are
allowed to move. Hence, a bone, where the extrinsic curvature
does not vanish, ends at the particle's vertex. 
\end{enumerate} 

Note that the dual to {\em the lattice of bones
and joints} is a triangular lattice; since every surface $\Sigma $ can
be triangulated this gauge induces no restriction in $\Sigma$'s
topology. 

Normal translation preserves this gauge, but it is {\em not} the only
evolution that preserves it. However, normal translations have the
advantage that they are generated by local expressions in terms of the
scalar variables. 

As seen above, orthonormal translations of vertices in the polygons
are generated by the deficit angles, which are local functions. Now we
show how local expressions for the generators of normal translations
of vertices at joints, bones or particles are derived. At a
joint $(J)$ three bones, with extrinsic curvatures $\eta _{1,2} , \eta
_{2,3} , \eta _{3,1}$, meet making angles $\alpha _1 , \alpha _2 ,
\alpha _3$; these angles are functions of the links' lengths $\alpha =
\alpha [l]$. Since we have only three bones going to a joint,  the
relation  ${\bf W(J)\approx 1}$ determines implicitly the angles
$\alpha (l)$ in terms of the extrinsic curvatures $\eta $ 

\nopagebreak[3]\begin{eqnarray} 
{\bf W}(J) [ \alpha _i , \eta _{ij} ] \approx {\bf 1} \Rightarrow  
\alpha _i[l's] \approx \alpha _i^* = \alpha _i^*[\eta _{ij}] \quad . 
\label{a(n)} 
\end{eqnarray} 
\noindent 
These relations between $\alpha 's$ and $\eta 's$ at a joint are {\em
local}. For a particle at the end of a bone we can produce a similar
expression of the deficit angle $\beta _p= \beta _p[l's]$ in terms of
the extrinsic curvature at the bone $\eta _p$ and the mass of the
particle $m(v)$ 

\nopagebreak[3]\begin{eqnarray} 
tr({\bf W}(p)[\beta _p , \eta _p])\approx 1 + 2 \cos(m(p) \nonumber \\ 
\Rightarrow   
\beta _p[l's] \approx \beta _p^*= \beta _p^*[\eta _p , m(p)] \quad .
\label{b(n,m)}
\end{eqnarray} 
\noindent 
't Hooft derived the explicit formulas (\ref{a(n)}), (\ref{b(n,m)}) for
the angles in terms of the extrinsic curvatures. He also proved that
the  generator of normal translations at $(v)$ $H(v)$  was simply the
deficit angle at $(v)$ 
\nopagebreak[3]\begin{eqnarray} 
H(v)&=& 2 \pi - {\sum _{i\to v}} \alpha _i^* \ \ 
\hbox{if there is no particle
at $(v)$} \nonumber \\ 
H(v)&=& \beta _p^*  \ \ 
\hbox{if there is a particle at $(v)$.}
\end{eqnarray} 
\noindent 
By (\ref{a(n)}), (\ref{b(n,m)}), we see that these are linear combinations
of the translation constraints. The local Hamiltonian constraints are
\nopagebreak[3]\begin{eqnarray} 
{\cal H}(v)&=& H(v) - (2 \pi - {\sum _{i\to v}} \alpha _i[l's]) 
\approx 0 \nonumber \ \ 
\hbox{if there is no particle at $(v)$} \nonumber \\ 
{\cal H}(v)&=& H(v) - \beta_p[l's] \approx 0 \ \ 
\hbox{if there is a particle at $(v)$.} \label{constr}
\end{eqnarray} 
\noindent 

Previously we saw that the generator of normal translations of vertices
in the interior of the domains is also the deficit angle; for calculational
purposes we can place the vertices at bones on the same footing as
joints. To do that, we consider one of the links connected to the
vertex to be a zero extrinsic curvature ``bone''. Thus, the generator of
normal translations of every vertex is the deficit angle at the
vertex, and the total deficit angle 
\nopagebreak[3]\begin{eqnarray} 
H = {\sum _v} H(v) \label{hamilt}
\end{eqnarray} 
\noindent 
can be regarded as a "Hamiltonian". 
Rather than in deriving 't Hooft's results \cite{'thooftCLASS,'thooftQUANT} 
by an alternative method,
we are interested in analyzing 't Hooft's quantization procedure using
the covariant lattice theory as a framework. We follow this procedure
because in the covariant theory the structure of the constrained
system is transparent. In particular, we have just found that the 
Hamiltonian (\ref{hamilt}) is constrained, by (\ref{constr}).
In the next section, we compare the results of different choices of
quantization for the lattice theory. 

\section{Four Quantizations}

\noindent {\bf 1. 't Hooft quantization.}  One assumes that the 
lattice is organized as a patchwork of extrinsically flat polygons, bones 
(polygon's boundaries)  and joints (vertices where bones meet) as 
described above; we will refer to these conditions 
hereafter as 't Hooft's gauge. The Hamiltonian constraints then fix the
values of the deficit angles. If one forms the sum of these Hamiltonians
over all lattice sites, the resulting function generates precisely the 
dynamical evolution which preserves 't Hooft's gauge. This suggests the
following quantum theory: one considers the phase space spanned by the link
lengths and boosts $l_{ij}, \eta_{ij}$, with the canonical brackets, and the 
Hamiltonian operator which corresponds from the sum of deficit angles at
the vertices. If the initial wave function has support only over $l_{ij}'s$ 
which respect 't Hooft's gauge, the same will hold at all times 
until the gauge crashes because a particle collides with a polygon's wall. 
Then one uses 't Hooft's prescription for "transitions" to get the initial 
conditions after the gauge's crash \cite{'thooftQUANT}. 

\

\noindent {\bf 2. Canonical Quantization in 't Hooft's Gauge}. 
One assumes a simplicial lattice 
described in terms of the scalar variables and Poisson brackets described 
above. The first class curvature constraints are complicated non-local 
functions of these variables; then one goes to 't Hooft's gauge and fixes 
all the translational gauge freedom in the direction non-normal to the 
polygons. After gauge fixing, one expects the Dirac brackets to be 
complicated non-local expressions. One then chooses
an internal time which labels the Cauchy surfaces, such as the length of a link
which connects two particles with a non-zero relative velocity. The momentum 
conjugate to this internal time is the corresponding boost parameter. The 
Hamiltonian constraint (sum of deficit angles) with the gauge conditions can
be solved for this momentum, leading to the physical Hamiltonian. Unlike the
previous case, this Hamiltonian (a boost parameter) is not a cyclic function
so there is no reason to expect that time would be quantized.

\

\noindent {\bf 3. Covariant Lattice Quantization}. One can quantize the 
lattice theory prior to fixing the frames. The link vectors become
operators in the Lie algebra -- in the Euclidian gravity case the algebra is
$so(3)$ and one has the prediction that the lattice links have a discrete 
spectrum, $l^2_{ij} = l(l+1)$. This leads to a picture which is reminiscent of
the original proposal of Ponzano and Regge, only with discrete space and 
continuous time.

\

\noindent {\bf 4. Quantization in the Polygon Representation.} One eliminates 
the pure gauge lattice structure down to the minimal lattice where all links
are either basis loops of the homotopy group or separate two particles. The
resulting link vectors have an $SO(2,1)$ ($SO(3)$) algebra as in the covariant
lattice, with the same remark with regard to the quantization of space.
An internal time parameter can be identified and the physical Hamiltonian
can be computed explicitly \cite{PhysrevD1}, leading to a quantization in the 
Schroedinger picture (with a continuous time parameter) \cite{BwedgeF,Quebec}.

\

\section{Discussion -- 't Hooft's Gauge in 3+1 Dimensions}

 We have shown that the representation of (2+1)--dimensional gravity in
terms of 't Hooft's polygons can be derived from the exact lattice formulation
of 2+1 gravity as the Chern-Simons theory of $ISO(2,1)$ connections. However,
the vertex Hamiltonians proposed by 't Hooft turn out to be first-class
constraints, which must be solved together with the gauge conditions that
specify a foliation by Cauchy surfaces that are patchworks of plat polygons. 

 If this gauge-fixing is carried out and the constraints are solved, one loses
the cyclic form of the Hamiltonian and the conclusion that time is quantized .
Vice-versa, if one chooses not to fix the gauge then this is not the correct
form of the Hamiltonian when the quantized polygons fluctuate away from 
't Hooft's gauge. 

 Besides the issue of quantization of time, we considered also the question
of whether space is quantized in the Euclidian version of the theory. Again, the
answer depends crucially on the choice of quantization. If the frames are
fixed prior to quantizing, one finds a continuous spectrum for the link
lengths -- but if the theory is quantized in a covariant manner then the link
lengths acquire the spectra of rotation operators.

 There is an extension of these results and of 't Hooft's gauge to
the lattice theory of topological gravity in 3+1 dimensions: an exact 
lattice theory has been proposed \cite{Lattice3+1} based on the 
similarity between the topological gravity and $B \wedge F$ theory 
\cite{BwedgeF}. In this theory the $SO(3,1)$ matrices
${\bf M}_{(ij)}$ define parallel-transport between neighboring
simplicial cells, 
and ${\bf E}_{(ij)}$ are the face bivectors in the local frames. One
can choose 
frames where the matrices are boosts which leave the separating face 
between 
two cells fixed.  Similar arguments to those followed here then show that the
product of Lorentz matrices around a bone of the lattice is a pure rotation in
a plane orthogonal to the bone; if one is considering the case of topological
gravity with no matter sources, this would be constrained to be weakly equal
to zero. The possibility of allowing a linear distribution of mass on the 
lattice 
bones and letting the deficit angle be non-zero is intriguing; perhaps such a 
construction can lead to a lattice theory for cosmic strings. 

This work was supported in part by Universidad Nacional Aut\'onoma de 
M\'exico (DGAPA), and grants
NSF-PHY-9423950, NSF-PHY-9396246, research funds of the Pennsylvania State
University, the Eberly Family research fund at PSU and the Alfred P.  Sloan
foundation. One of us (JZ) would like to thank Alejandro Corichi and Monica 
Pierri for helpful discussions, and Mary-Ann Hall for her editorial help.

\end{document}